\documentclass[authoryear,final,12pt,twocolumn]{elsarticle}
\usepackage{amssymb}
\journal{Planetary and Space Science}
\begin{document}
\begin{frontmatter}
\title{Symbolic analysis of slow solar wind data using rank order statistics}
\author{Vinita Suyal}
\author{Awadhesh Prasad}
\author{Harinder P. Singh\corref{cor3}}
\ead{hpsingh@physics.du.ac.in}
\cortext[cor]{Corresponding author}
\address{Department of Physics and Astrophysics,
University of Delhi,
Delhi - 110007, India.}
\begin{abstract}
We analyze time series data of the fluctuations of slow solar wind velocity using rank order statistics. We selected a total of $18$ datasets measured by the Helios spacecraft at a distance of $0.32$ AU from the sun in the inner heliosphere. The datasets correspond to the years $1975-1982$ and cover the end of the solar activity cycle $20$ to the middle of the activity cycle $21$. We first apply rank order statistics to time series from known nonlinear systems and then extend the analysis to the solar wind data. We find that the underlying dynamics governing the solar wind velocity remains almost unchanged during an activity cycle. However, during a solar activity cycle the fluctuations in the slow solar wind time series increase just before the maximum of the activity cycle.
\end{abstract}
\begin{keyword}
Solar wind plasma; Solar activity; Symbolic analysis.
\end{keyword}
\end{frontmatter}
\section{Introduction}
\label{intro}
Solar wind is a stream of charged particles, mostly protons and electrons, that are ejected from the upper atmosphere of the sun. \cite{park58}, who first coined the term `solar wind', showed that the solar wind is in fact the outer coronal atmosphere escaping supersonically into the interstellar space. The solar wind is primarily divided into two components, a fast wind with velocities $\sim$900 km/s and a slow wind with velocities $\sim$300 km/s. These two components differ in their origin and characteristics which include flow speed, proton density, temperature, and helium content \citep{feld05,schw06}. The fast solar wind originates from coronal holes, where the magnetic field lines are open, while slow wind is associated with open magnetic field and is highly structured and highly variable \citep{krie73,woo97}.

Observations of solar wind velocity distributions made by different spacecrafts have been analyzed in considerable detail. Helios spacecraft probed the heliosphere as close as $0.3$ AU from the sun and these observations have been analyzed using linear and nonlinear time series analysis tools \citep {mace97,mace98a,mace98b,mace00,reda01,gupt08}. \cite{mace97} argued that the inner heliosphere is a low dimensional chaotic attractor. \cite{mace98a} studied a single time series using moving average, singular-value decomposition and surrogate data analysis to characterize the nonlinear behavior of the data. \cite{mace00} have shown that the Kolmogorov entropy is positive and finite, as it holds for a chaotic system. \cite{reda01} extended the study by using a nonlinear filter to obtain the noise-reduced data and more reliable estimate of largest Lyapunov exponent and Kolmogorov entropy. \cite{gupt08} analyzed several Helios time series to understand the local dynamics of fluctuations in the slow solar wind velocity. They suggest the inherent changes in the dynamics over the solar cycle using the variance in largest Lyapunov exponent for each time series.\\
\begin{table}[t]
\caption{The solar wind velocity data measured by the Helios spacecraft in the years 1975 to 1981 with Carrington rotation(CR); Initial time ($T_{i}$); Final time($T_{f}$); Data points($N$). Sunspot numbers are from http://sidc.oma.be/sunspot-data/.}
\vspace{.5cm}
\resizebox{8cm}{!} {
\begin{tabular}{|ccccccc|}
\hline
S.No.&CR&Year&$T_{i}$&$T_{f}$&$N$&Sunspot\\
&&&(day:h:min) &(day:h:min)& &number\\
\hline
1&1625&1975&069:21:36&071:17:42&3515&18\\
2&1625&1975&077:00:45&078:20:52&2938&22\\
3&1632&1975&260:00:35&261:20:39&2979&14\\
4&1632&1975&267:03:52&268:23:56&3083&0\\
5&1639&1976&085:04:28&087:00:45&2245&38\\
6&1639&1976&092:07:20&094:03:10&1843&23\\
7&1646&1976&282:11:52&284:06:39&1188&7\\
8&1653&1977&099:11:25&101:07:15&2741&0\\
9&1653&1977&106:13:51&108:09:57&3376&31\\
10&1660&1977&296:17:34&298:13:48&1085&30\\
11&1667&1978&114:19:36&116:14:33&2393&106\\
12&1667&1978&121:21:18&123:17:31&1911&89\\
13&1681&1979&137:05:09&139:01:23&1894&148\\
14&1695&1980&145:10:38&147:07:04&2218&229\\
15&1695&1980&152:12:36&154:09:02&2559&152\\
16&1709&1981&159:17:43&161:14:02&3174&58\\
17&1709&1981&167:03:16&168:16:26&2562&119\\
18&1723&1982&175:00:30&176:21:15&1361&112\\
\hline
\end{tabular}
}
\label{tab}
\end{table}

In this work, we study how the dynamical system governing the slow solar wind velocity  behaves at different phases of the solar activity cycle. We apply the method of rank order statistics of symbolic sequence \citep{yang03} to investigate the solar wind time series data. In the next Section, we describe the datasets and their pre-processing. In Section \ref{symb} we describe the rank order statistics of symbolic sequence technique. In Section \ref{resu} we apply symbolic analysis technique to time series obtained from known nonlinear dynamical systems like Lorenz, R\"ossler, and Chua oscillators and extend the analysis to solar wind data. In the last Section we give conclusions of our study.\\
\begin{figure*}
\centering
\includegraphics[width=10.0cm]{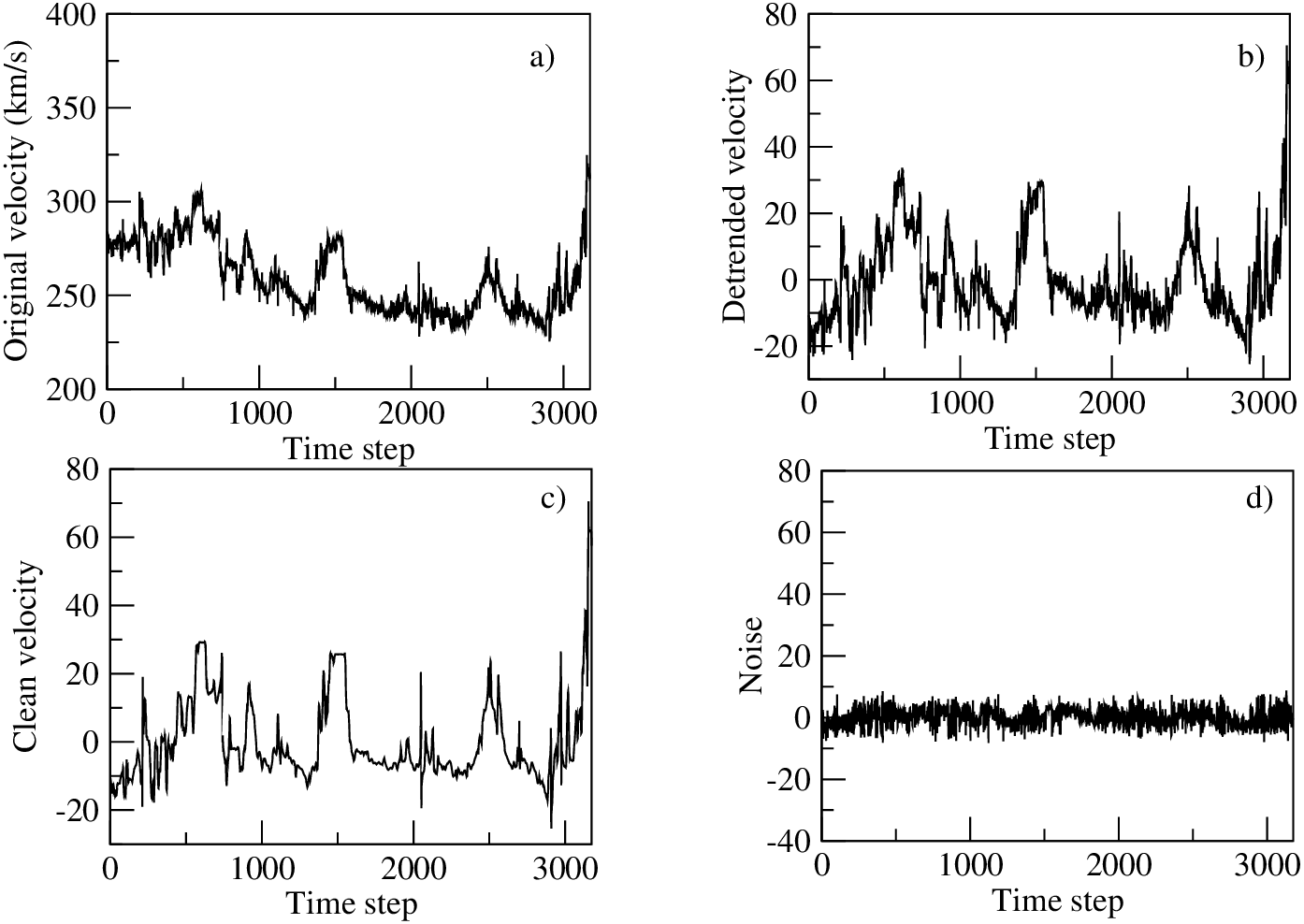}
\caption{Pre-processing of an example data set (time series number $16$ in table \ref{tab}) (a) the original data, (b) the detrended data, (c) the filtered data after nonlinear noise reduction, and (d) the noise removed by nonlinear filter.}
\label{fig1}
\end{figure*}
\section{Helios data and noise reduction}
We analyze time series data of radial velocity ($v$) measured by the Helios spacecraft at $0.32$ AU in the years $1975$ to $1981$, corresponding to parts of the solar activity cycles $20$ and $21$ (June, $1976$ to September, $1986$). This data is obtained from http://sprg.ssl.berkeley.edu/impact/data\_\\browser\_helios.html. Although several more time series were available, only those were selected that satisfied the condition of sufficient ($>1000$) data points. Thus, $18$ time series were extracted from the Helios database with the sampling time for each time series being $40.5$s. The details of the time series used in the analysis are given in Table 1. In addition, Table 1 (last column) contains the corresponding sunspot numbers for each time series obtained from http://sidc.oma.be/sunspot-data/.\\

In Fig. \ref{fig1}a , we have shown a solar wind time series of $3174$ data points recorded in the year $1981$. The sequence shows  a general decreasing trend. Therefore, before a detailed analysis is carried out, linear and quadratic trends ($v=at^2+bt+c$, with $t$ being a fraction of the total sample) were subtracted from the raw data \citep{mace97,gupt08}. For the data shown in Fig.\ref{fig1}a, parameters $a$, $b$, and $c$ are $0.15$, $-1.71$, and $294.77$ respectively. Fig. \ref{fig1}b shows the resulting detrended time series for the data.\\

Solar wind data are believed to be a output of a dynamical process upon which the noise has been superimposed \citep{mace98a, mace00}. Following \cite{mace00} and \cite{ gupt08} we use a nonlinear filter \citep{schr93a} to clean the signal. This filter averages in the embedding space of a chosen dimension $2l+1$ and a defined neighborhood of size $\epsilon$, about $2$-$3$ times of the estimated initial noise level \citep{schr93b}.\\
\begin{figure*}
\centering
\includegraphics[height=7cm,width=8cm]{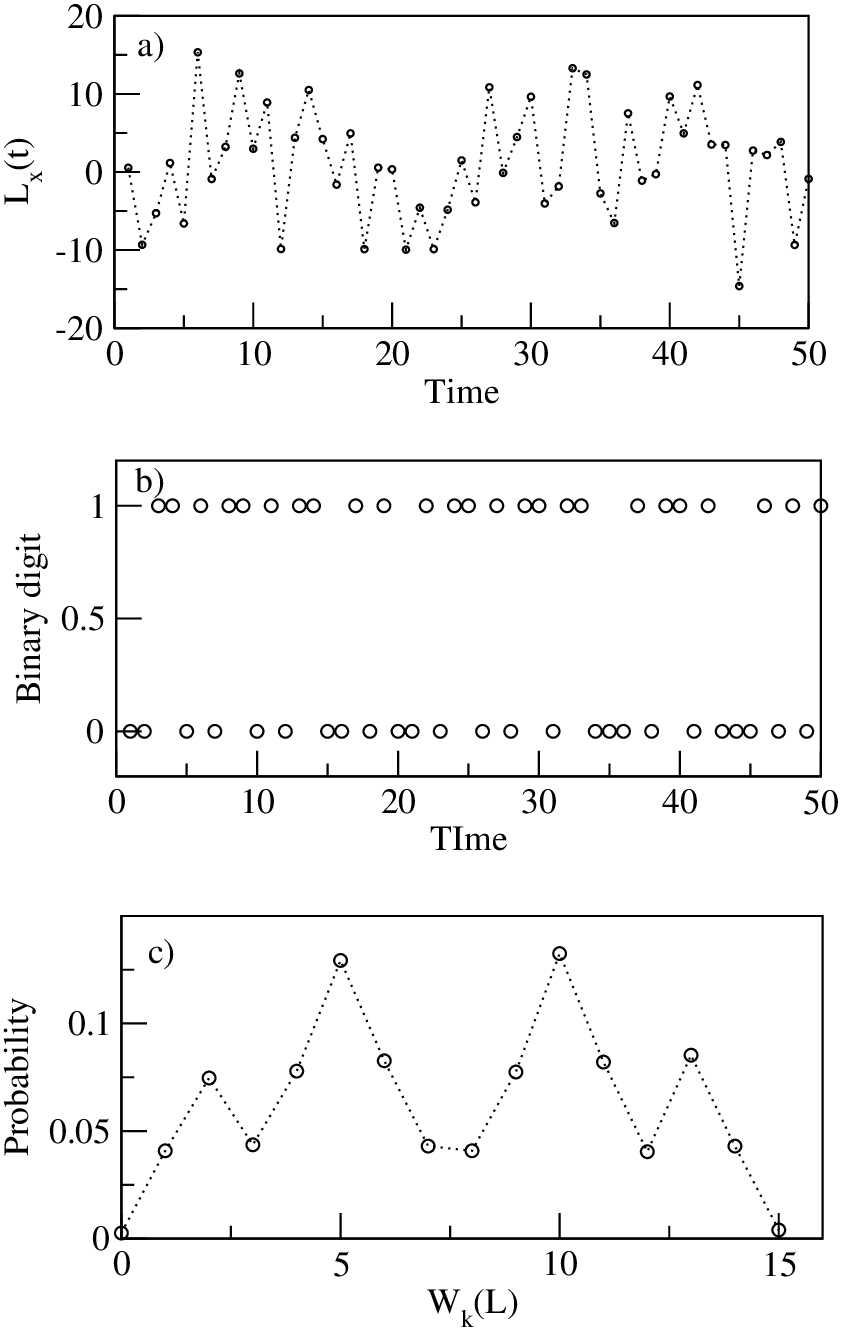}
\caption{(a) Time series (x component) of the Lorenz oscillator, Eq. (5), (b) binary digit corresponding to the time series, binary word corresponds to $t=1$ is $(2^3.0+2^2.0+2^1.1+2^0.1=3)$, (c) probability distribution of every 4-bit word. The word index ranges from $0$ to $2^L -1$ for $L$- bit words (complete time series).}
\label{fig2}
\end{figure*}
Let $v_i, i=1,2,...,n$ be the detrended signal. We construct vectors $V_i=(v_{i-l},...,v_{i+l})$ in this $2l+1$ embedding space. We then replace the data point $v_i$ by its mean value in the neighborhood of size $\epsilon$,
\begin{equation}
\centering{v_i^c=\frac{1}{N_i}\sum_{|V_i-V_j|<\epsilon}v_j,}
\end{equation}
Where $v_i^c$ denote the clean value and $N_i$ is the number of neighbors within $\epsilon$, i.e., $|V_i-V_j|<\epsilon$.
This process is now iterated taking decreasing $\epsilon$ until no neighbors are found. We performed two iterations taking the parameters: $l=3, \epsilon =0.12$ and $l=6, \epsilon =0.06$ \citep{schr93a}. The noise (that is, the difference between the detrended data in Fig. \ref{fig1}b and the filtered data in Fig. \ref{fig1}c) is shown in Fig. \ref{fig1}d.
\label{noise}
\begin{figure*}
\centering
\includegraphics[width=13.0cm,height=6.5cm]{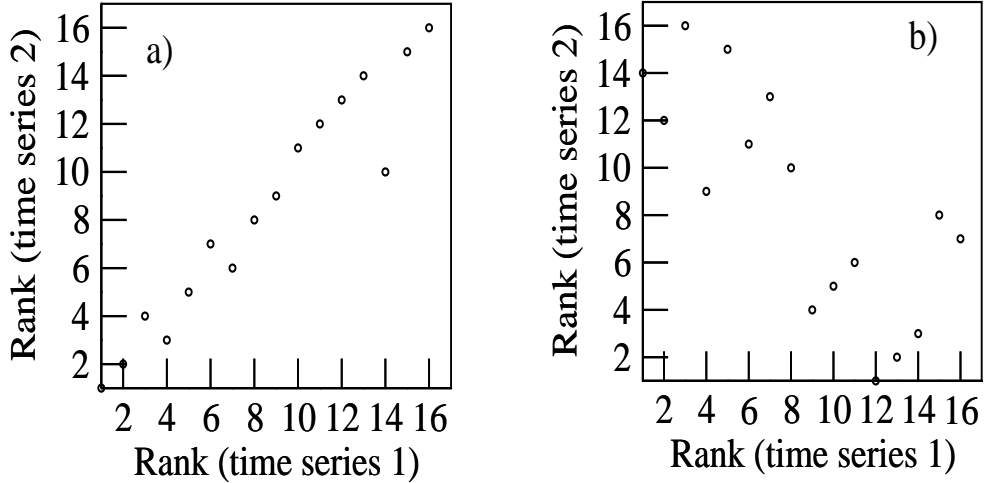}
\caption{Rank order comparison of two time series from (a) same system ($x$ component of Lorenz oscillator at different time), (b) different system ($x$ component of Lorenz oscillator and $x$ component of R\"ossler oscillator).}
\label{fig3}
\end{figure*}
\section{Symbolic Analysis}
\label{symb}
Time series that originate from dynamical systems having complex nonlinear interactions may look very much like noise \citep{kant04}. Dynamical system behind such intrinsically noisy dataset can be simplified by mapping it to the binary sequences, where the increase and decrease of fluctuations are denoted by $1$ and $0$ \citep{kurt95}.\\

Let $x_i, i=1,2,...,n$ be the time series data. A part ($50$ data points) of the time series (here $x$ component of Lorenz oscillator \citep{lore63}) is shown in Fig. \ref{fig2}a.  Following \cite{yang03}, we classify each pair of successive data into one of the two states that represents a decrease in $x$, or an increase in $x$. These two states are mapped to the symbols $0$ and $1$, respectively:
\begin{eqnarray}
\nonumber
S_n &=& 0  \quad {x_n \le  x_{n-1}}\\
 &=& 1  \quad {x_n > x_{n-1}}
\end{eqnarray}
We take $S_1=0$. This binary sequence represents a unique pattern of fluctuations in a given time series. Fig. \ref{fig2}b shows
 sequence of binary digits for the part of the data shown in Fig. \ref{fig2}a. The resulting symbolic series is then partitioned into short sequences of given length L. Every short sequence is then uniquely identified by a unique word (integer),
\begin{equation} 
\centering{w_L(k)=\sum_{i=1}^{L} 2^{L-i}S_{i+k-1}.}
\end{equation}
We find the probabilities of different words (Fig. \ref{fig2}c) in the complete time series, and then sort them according to decreasing frequency. The resulting rank-frequency distribution, therefore, represents the statistical hierarchy of symbolic words of the time series. We compare the rank number of each $L$ bit word for two different time series. In Fig. \ref{fig3} we plot the rank number of two time series against each other. Fig. \ref{fig3}a shows the rank order comparison for time series of $x$ component of chaotic solution of Lorenz oscillator at two different time regimes. In Fig. \ref{fig3}b rank order comparison for time series of $x$ component of Lorenz oscillator and time series of $x$ component of R\"ossler oscillator in the chaotic regime is shown. If the two time series are similar (from same dynamical system), their rank order points will be located near the diagonal line (Fig. \ref{fig3}a). If the two time series are from different dynamical systems, their rank points will be more scattered (Fig. \ref{fig3}b). We define a parameter weighted distance for $L$ bit sequence ($D_L$) between the two time series as:

\begin{equation}
\tiny{D_L(x,y)=\frac {\sum_{k=1}^{2^L} 
\mid R_x(w_k)-R_y(w_k)\mid  p_x(w_k)p_y(w_k)}
{(2^L-1)\sum_{k=1}^{2^L}p_x(w_k)p_y(w_k)}}.
\label{eqn4}
\end{equation}
\begin{figure*}[ht]
\begin{minipage}[b]{0.5\linewidth}
\centering
\includegraphics[width=6.5cm,height=5cm]{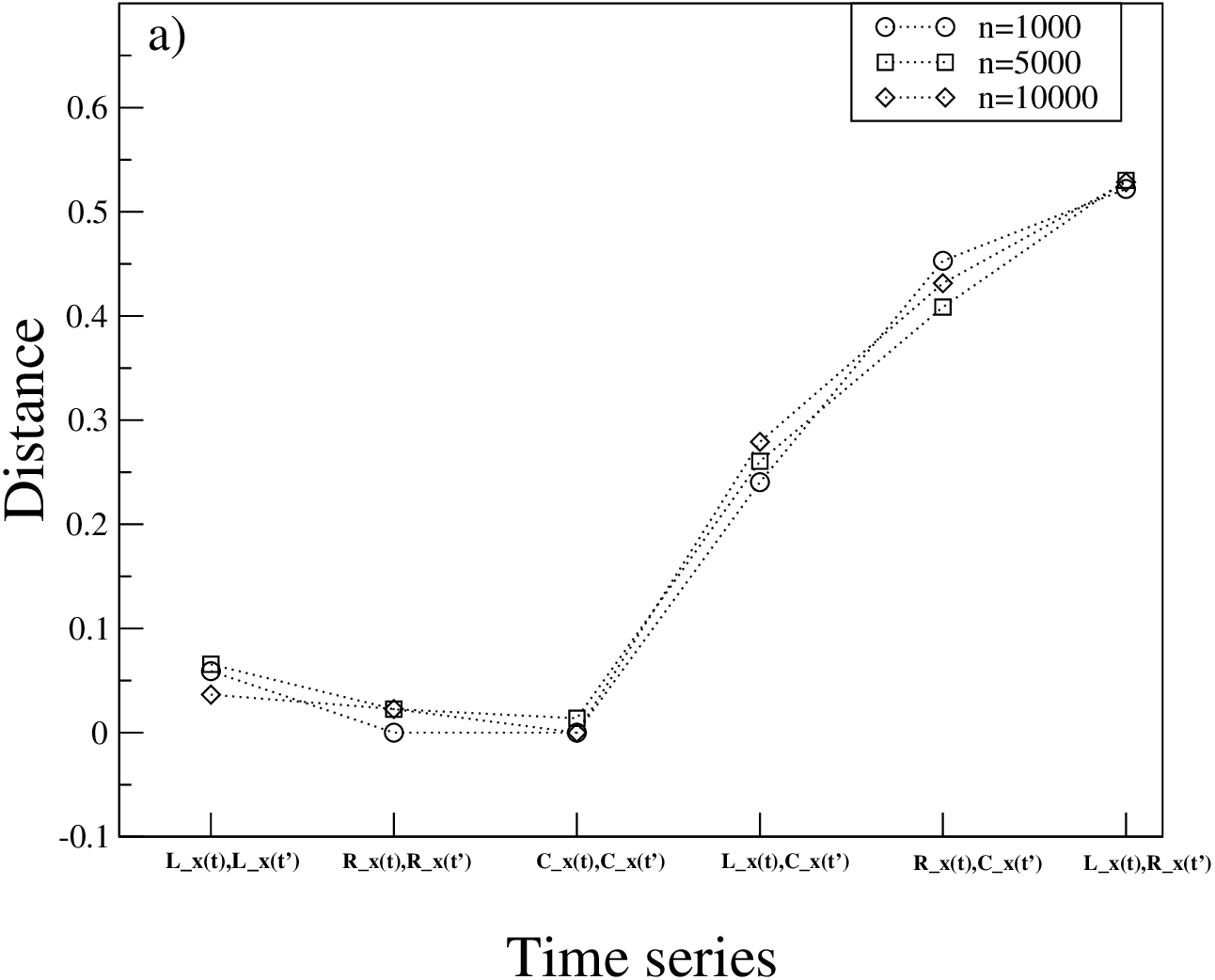}
\end{minipage}
\hspace{.1cm}
\begin{minipage}[b]{0.5\linewidth}
\centering
\includegraphics[width=6.5cm,height=5cm]{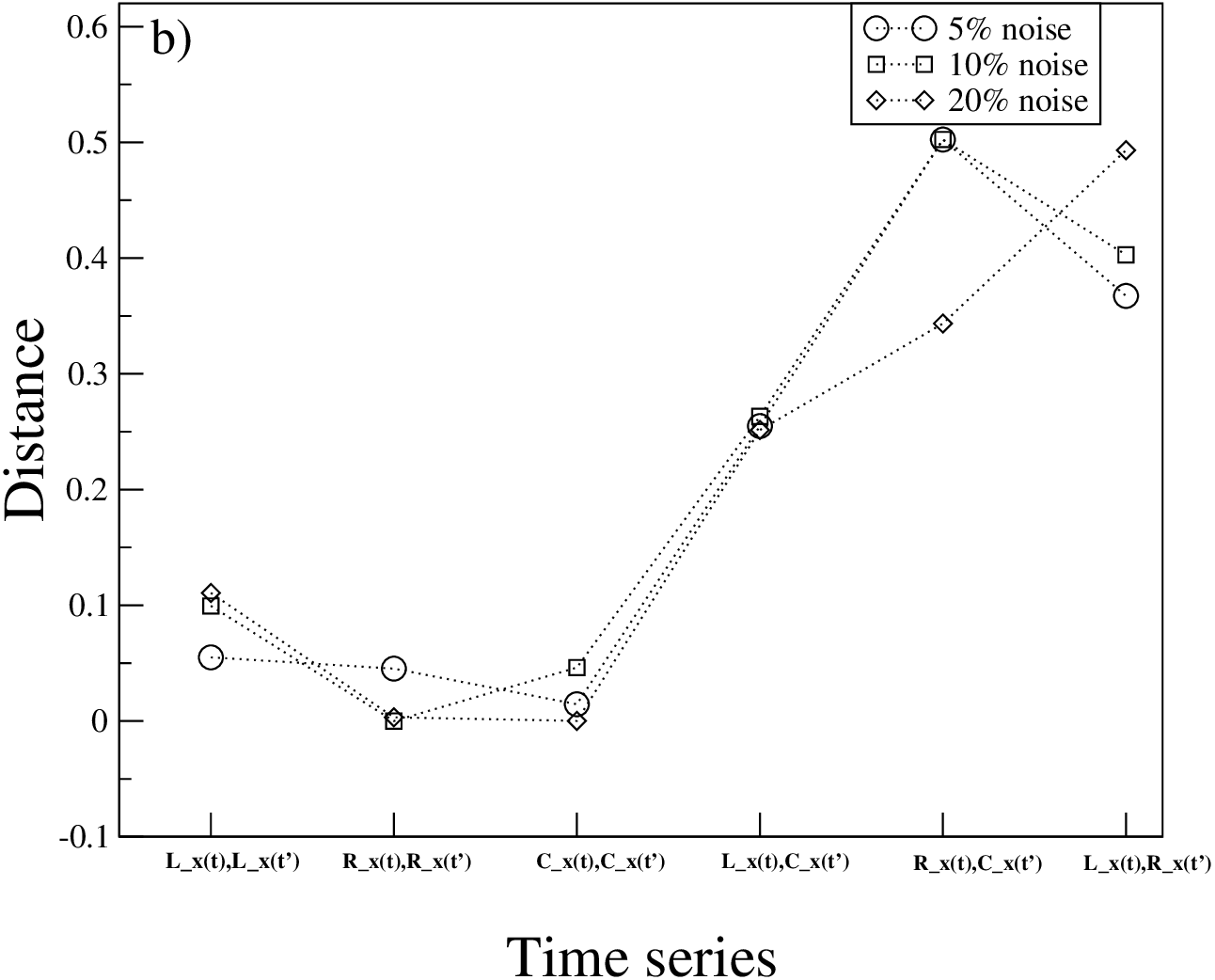}
\end{minipage}
\caption{Distances $D_L$ for different combinations of time series obtained from nonlinear dynamical systems (a) for time series of length $1000$, $5000$, and $10000$, (b) with $5$\%, $10$\%, and $20$\% noise added to the time series of  $1000$ data points.}
\label{fig4}
\end{figure*}
\begin{figure*}
\centering
\includegraphics[width=14cm,height=6cm]{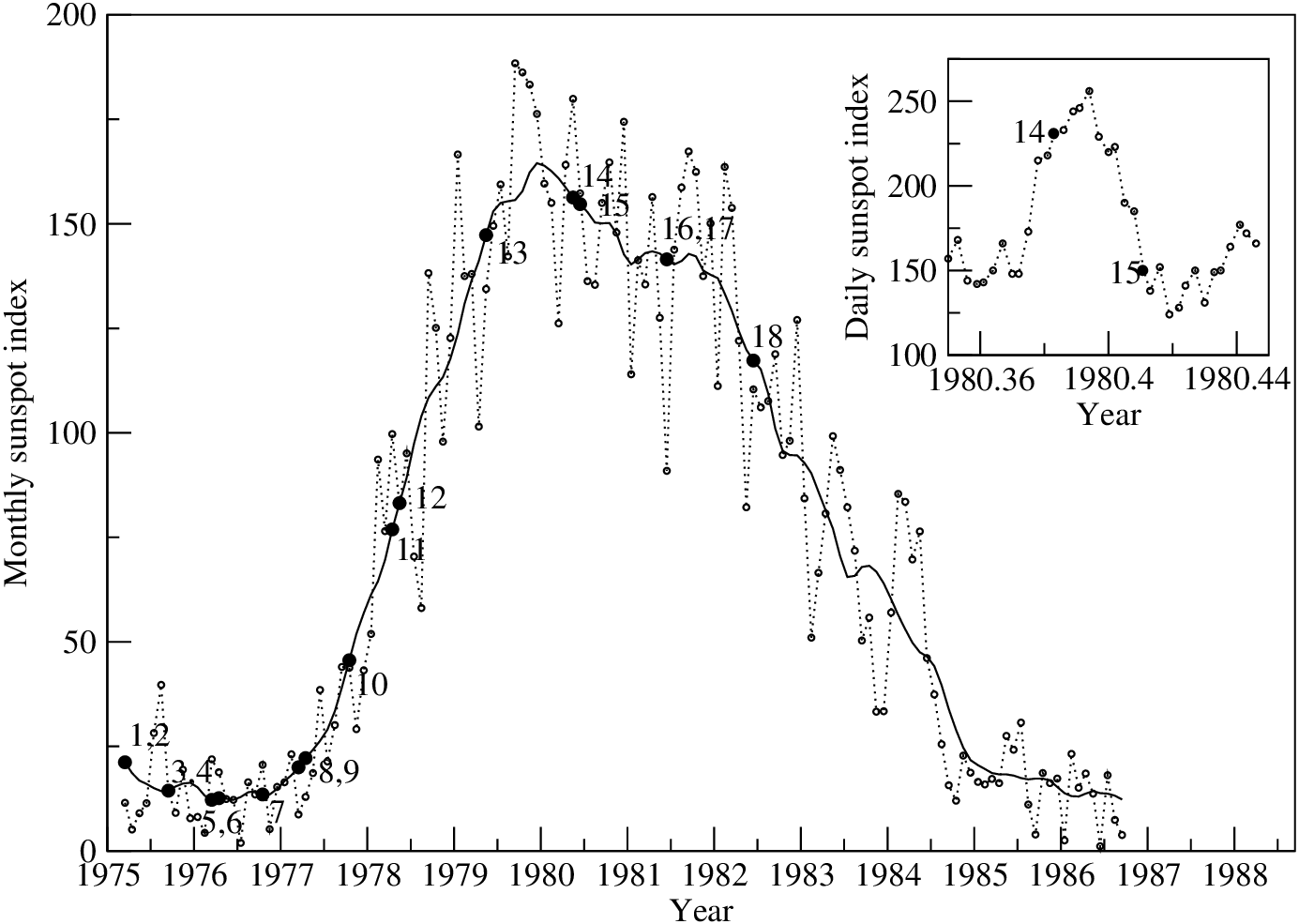}
\caption{Monthly averaged sunspot number (broken line) and smoothed monthly sunspot number (solid line) from March $1976$ to September $1986$. Initial positions of solar wind data sets are shown on the smooth curve. In the inset daily sunspot index and position of data sets $14$ and $15$ are shown.}
\label{fig5}
\end{figure*}\\
\\
\begin{figure*}
\begin{minipage}[b]{0.5\linewidth}
\centering
\includegraphics[width=7.5cm,height=7.5cm]{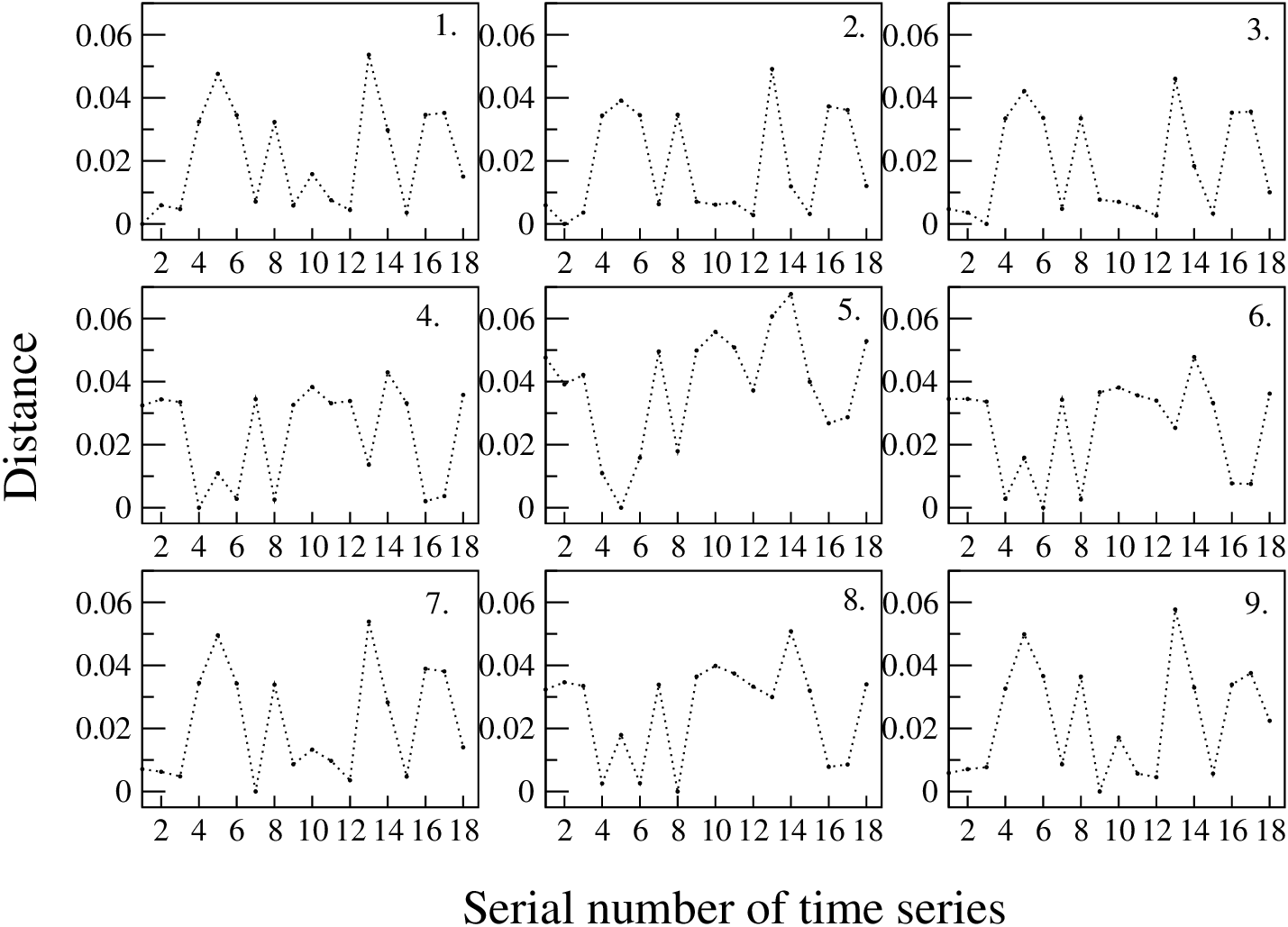}
\end{minipage}
\hspace{.1cm}
\begin{minipage}[b]{0.5\linewidth}
\centering
\includegraphics[width=7.7cm,height=7.5cm]{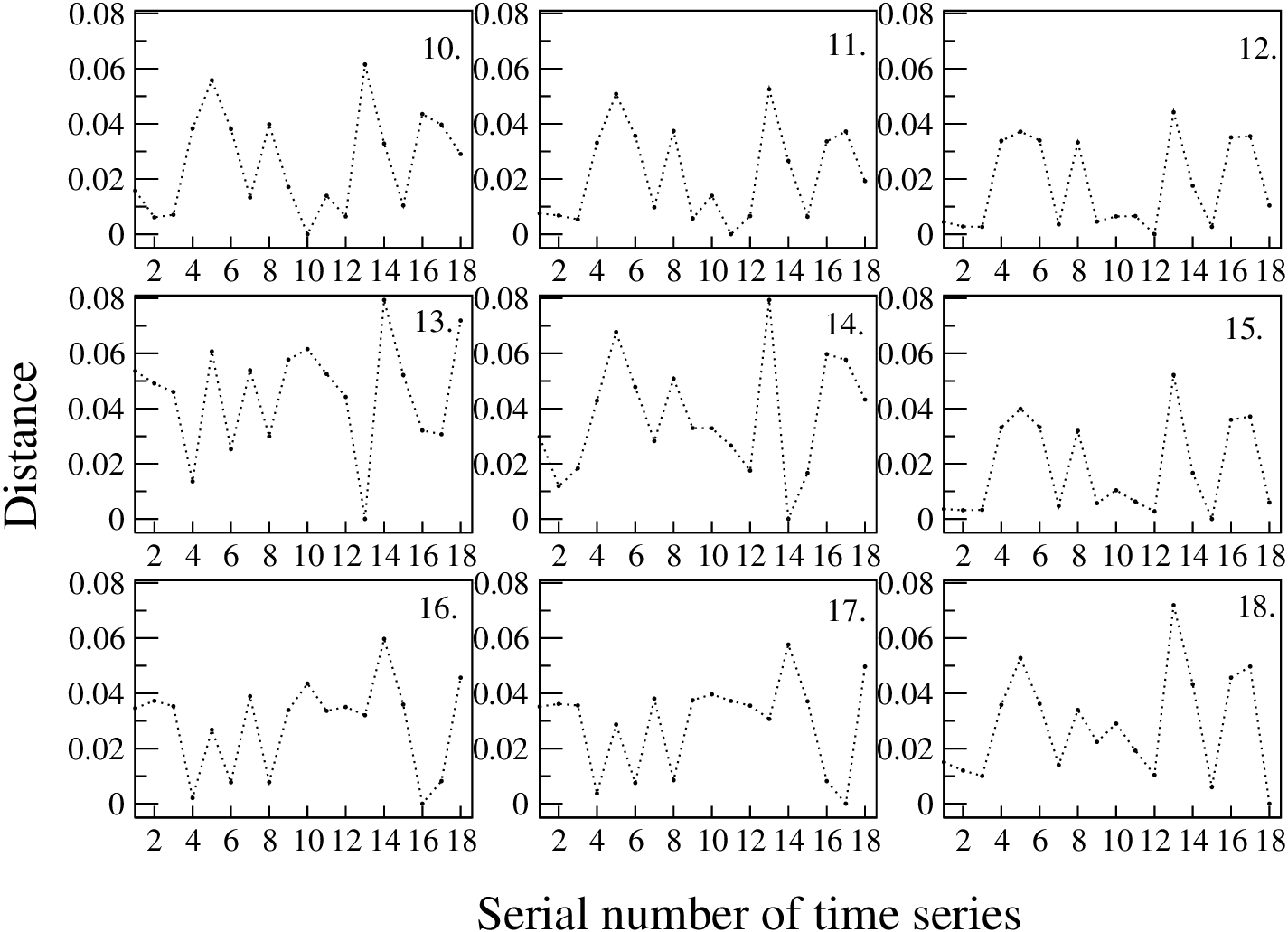}
\end{minipage}
\caption{Distances $D_L$ of a time series from others. For example, panel $1$ (top left) shows the distance between time series $1$ and all $18$ time series. The distance between same two time series is zero.}
\label{fig6}
\end{figure*}
\begin{figure*}
\centering
\includegraphics[width=12.0cm,height=5.5cm]{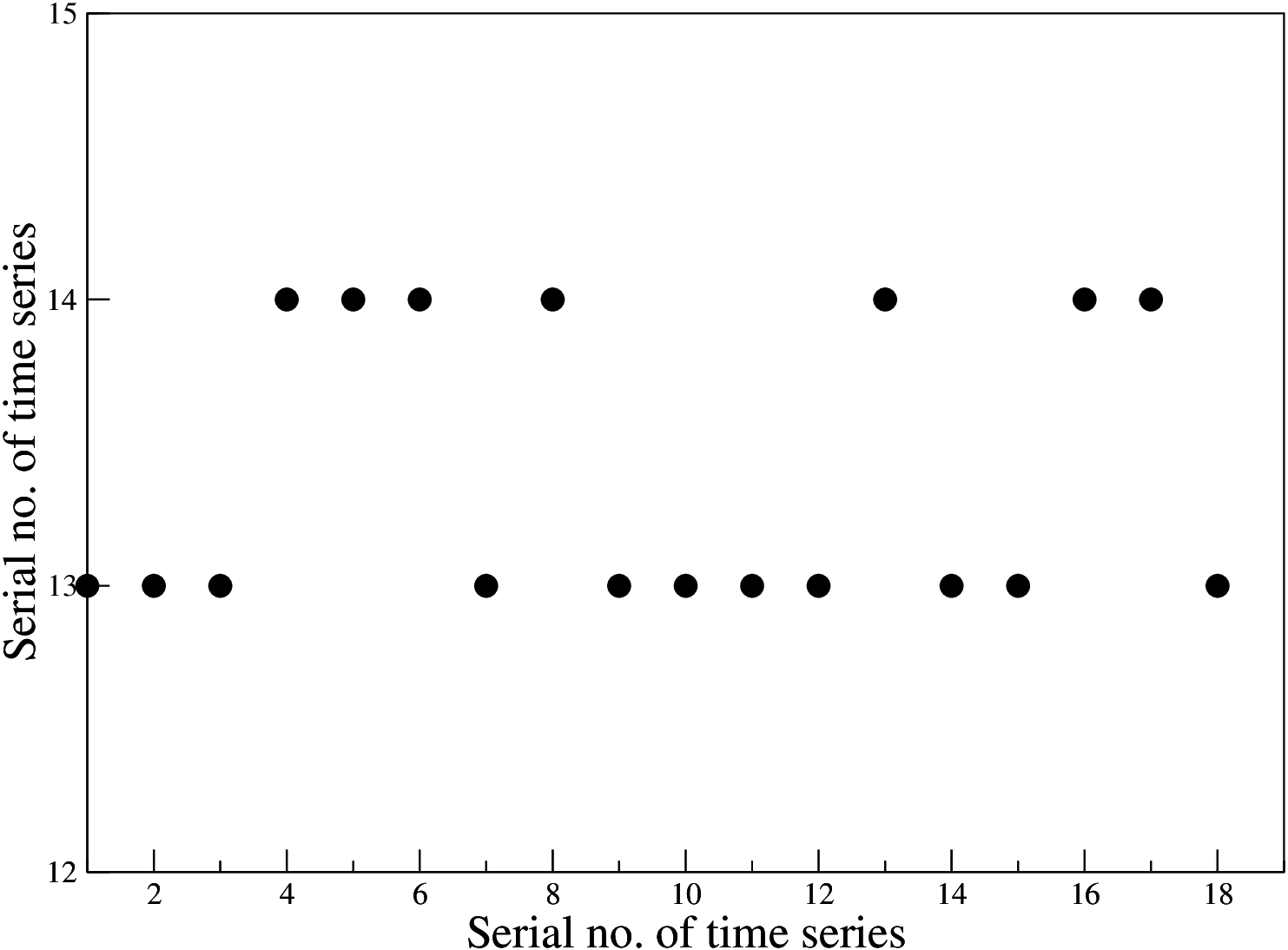}
\caption{Serial number of time series vs. most distant time series serial number.}
\label{fig7}
\end{figure*}
Here x and y are two time series, $p(w_k)$ and $R(w_k)$ represent the probability and the rank of a specific word ($w_k$) in the time series respectively. Value of $D_L$ lies between $0$ and $1$. Greater distance indicates less similarity between the two time series and vice-versa. The distance between same two time series is zero.
\section{Results and Discussion}
\label{resu}
\subsection{Rank order statistical analysis of nonlinear models}
We take three systems (Lorenz, R\"ossler, and Chua) exhibiting chaotic motions. Lorenz system is give by \citep{lore63}:
\begin{eqnarray}
\dot x &=& \sigma(y-x),\nonumber\\
\dot y &=& x(\rho -z)-y,\nonumber\\
\dot z &=& xy-\beta z.
\end{eqnarray}
We choose parameters $\sigma=10, \rho =28$, and $\beta=8/3$.\\
R\"ossler system is given by \citep{ross76}:
\begin{eqnarray}
\dot x & =& -y-z ,\nonumber\\
\dot y &=& x+ay ,\nonumber\\
\dot z& =& b+z(x-c).
\end{eqnarray}
Parameters are fixed at $a=0.1, b=0.1$, and $c=18$.\\
Chua system is given by \citep{chua93}:
\begin{eqnarray}
\dot x &=& c_1(y-x-p(x)),\nonumber \\
\dot y &=& c_2(x-y+z),\nonumber\\
\dot z &=& -c_3y,
\end{eqnarray}
where $p(x)$ is defined as,
$p(x)=m_1x+
\left((m_0-m_1)(\mid x+1\mid 
- \mid x-1 \mid)\right)/2$.
We take $c_1 = 15.6$, $c_2 = 1,
 m_0=-8/7,~m_1 = 5/7$ and select
the parameter $c_3=33$ such that the motion will be a single-scroll type.\\

We take two time series from each system at different time intervals. Hence we have $6$ time series in all.
They are denoted as $L_x(t)$ and $L_x(t^{'})$ for the Lorenz oscillator, $R_x(t)$, $R_x(t^{'})$ for the R\"ossler oscillator, and $C_x(t)$ and $C_x(t^{'})$ for the Chua Oscillator. We use rank order statistics on these time series data by keeping $L=4$. Fig. \ref{fig4}a shows distances ($D_L)$ for different combinations of these time series of length $1000$, $5000$ and $10000$. It is clear from Fig. \ref{fig4}a that time series from the same system have less distance as compared to time series from different systems. Fig. \ref{fig4}a also indicates that $1000$ data points are sufficient for this analysis.
Similar results are obtained when we add noise to the original time series. Fig. \ref{fig4}b shows the result when we add $5$\%, $10$\%, and $20$\% noise to the original time series of $1000$ data points. Therefore, we can say that this rank order statistics is quite robust against the presence of high level external noise.
\subsection{Rank order statistical analysis of solar wind}
Solar wind data have positive largest Lyapunov exponent \citep{pavl92,mace01,gupt08}, positive Kolmogorov entropy \citep{mace00,mace01}, and fractal dimension \citep{mace97}. This leads to the conclusion that wind dynamics are driven by the complex nonlinear interactions. We use rank order statistics with $L=4$ and time series of $1085$ data points for the analysis.
Fig. \ref{fig5} shows the solar activity from March $1975$ to September $1986$. It shows monthly averaged sunspot index and smoothed monthly sunspot index. In the inset daily sunspot index around the maximum of solar activity cycle is shown. We have shown starting time of our analyzed time series on the activity cycle. Time series $13$, $14$, and $15$ are located around the maximum of solar activity cycle $21$. Sunspot index corresponding to them are $148$, $229$, and $152$ respectively i.e., sunspot index increases while going from time series $13$ to $14$ and decreases on going  from time series $14$ to $15$.

Fig. \ref{fig6} shows the inter-time series distance of solar wind velocity as given by Eq. (\ref{eqn4}). Behavior of inter-time series distance remains same for all the time series, but the distance is maximum with time series $13$ (for time series $1,2,3,7,9,10,11,12,14,15,18$) and with time series $14$ (for time series $4,5,6,8,13,16,17$), i.e., only two time series $13$ and $14$  show least correlation with rest of the time series. Fig. \ref{fig7} shows the most distant time series from a particular time series.
 Positions of these time series are around the maximum of the solar activity cycle but towards the increasing phase only. Time series $15$ is also near the solar maximum but it is in the decreasing phase. This could be a reason for its less distance to the other time series as compared to the time series $14$.

\section{Conclusions}
\label{conc}
We used rank order statistics to analyze the time series obtained from known nonlinear systems. This method is an effective tool for analyzing short time series even in the presence of high level of external noise and is applied to $18$ solar wind time series obtained by the Helios spacecraft for the years $1975-1982$. Nonlinear filter is used to reduce the noise present in the data. We found that all the time series have least correlation with data sets $13$ and $14$, which are located near the maximum of the solar cycle. Although data set $15$ is also located very close to the data set $14$ (on the solar activity cycle), it shows significant correlation. One possible reason for this may be that data set $15$ is located in the decreasing phase of the solar activity cycle, while data sets $13$ and $14$ are located in the increasing phase of the solar activity cycle. This suggests the presence of larger fluctuations in the solar wind velocity data corresponding to the time just before the maximum of the activity cycle.
\section*{Acknowledgements}
The authors thank Prof. Rainer Schwenn and Prof. Eckart Marsch, Max Planck Institute for Solar System Research, Lindau, Germany, for the Helios data. VS thanks CSIR, India for a Senior Research Fellowship and  AP acknowledges DST, India for financial support.

\end{document}